\documentclass[aps,floats,prb,showpacs,twocolumn]{revtex4}

\newcommand{\beq}{\begin{eqnarray}} \newcommand{\eeq}{\end{eqnarray}}

\usepackage{amsfonts,amsmath} \usepackage{bm} \usepackage{dcolumn}
\usepackage{epsfig} \usepackage{latexsym}
\usepackage{graphicx}% Include figure files
\begin{document}

\title{Weak Localization in Metallic Granular Media}

\author{Ya.M. Blanter$^{1,2}$, V.M. Vinokur$^3$, and L.I. Glazman$^{4}$}

\affiliation{$^1$Kavli Institute of Nanoscience, Delft University
of Technology, 2628CJ Delft, the Netherlands\\
$^2$Braun Center for Submicron Research and Department of Condensed Matter
Physics, Weizmann Institute of Science, Rehovot 76100, Israel\\
$^3$Materials Science Division,  Argonne National Laboratory, Argonne,
IL 60439, USA
\\
$^4$W.I. Fine Theoretical Physics Institute, University of
Minnesota, Minneapolis, MN55455, USA}

\date{January 10, 2006}


\begin{abstract}
  We investigate the interference correction to the conductivity of a
  medium consisting of metallic grains connected by tunnel junctions.
  Tunneling conductance between the grains, $e^2g_{\rm T}/\pi\hbar$, is
  assumed to be large, $g_{\rm T}\gg 1$. We demonstrate that the weak
  localization correction to conductivity exhibits a crossover at
  temperature $T\sim g^2_{\rm T}\delta$, where $\delta$ is the mean
  level spacing in a single grain. At the crossover, the phase
  relaxation time determined by the electron-electron interaction
  becomes of the order of the dwell time of an electron in a grain.
  Below the crossover temperature, the granular array behaves as a
  continuous medium, while above the crossover the weak localization
  effect is largely a single-junction phenomenon. We elucidate the
  signatures of the granular structure in the temperature and magnetic
  field dependence of the weak localization correction.
\end{abstract}

\maketitle

\section{Introduction}

Quantum effects in conduction of two-dimensional disordered electron
systems draw attention of both experimentalists and theorists for
decades. The interest is motivated in part by the interplay between
several fundamental physical phenomena, such as quantum interference,
localization, superconductivity, and single-electron tunneling
occurring in these systems. The interplay affects the properties of
normal conductors~\cite{Gershenson,Beenakker,Sarachik}, nominally
superconducting films~\cite{Goldman,Dynes}, and arrays of
junctions~\cite{Fazio}. Quantum effects become increasingly important
at sheet conductances decreasing towards the fundamental quantum value
of $G_Q\equiv e^2/\pi\hbar$. The interpretation of some of the most
intriguing data, however, may depend on whether the investigated
conductors are homogeneous or granular. While this question has a
definite answer in the case of an array~\cite{Fazio} of
  lythographically defined junctions, it is less clear for nominally
homogeneous deposited metallic films~\cite{Goldman,Dynes} or electron
gases in semiconductor heterostructures~\cite{Sarachik}. Checking the
samples homogeneity traditionally involves application of auxiliary
techniques, such as local probe spectroscopy~\cite{Dynes,Yacoby}.

We demonstrate that information about the granularity of a conductor
is contained in the temperature and magnetic field dependence of the
weak localization (WL) correction to the conductivity. The granular
structure of a conductor affects the correction even at high film
conductivity, $\sigma_0\gg G_Q$. While being universal at the lowest
temperatures and magnetic fields, the WL correction becomes
structure-dependent at higher values of field and temperature. The
corresponding crossover temperature is of the order of
$(\sigma_0/G_Q)^2\delta$, where the mean-level spacing $\delta$ in a
single grain is inversely proportional to the grain volume. The field
dependence of the WL correction at low temperatures exhibits two
crossovers. These are associated with a significant change in structure of
closed electron trajectories, allowing for phase-coherent electron motion.

The WL correction in a homogeneous medium originates from the
quantum interference of electrons moving along self-intersecting
trajectories~\cite{Gershenson} and is proportional to the return
probability of an electron diffusing in a disordered medium. In
one- or two-dimensional conductors this probability diverges due
to the contribution coming from long trajectories. For a fully
coherent electron propagation, this divergence would lead to a
divergent WL correction. Finite phase relaxation time $\tau_\phi$
makes sufficiently long trajectories ineffective for the
interference, and limits the correction. In a two-dimensional
conductor, the WL correction to conductivity is
$\delta\sigma=-(G_Q/2\pi)\ln (\tau_\phi/\tau)$, where $\tau$ is the
electron momentum relaxation time. There are various mechanisms of
the electron phase relaxation, some of them being
material-specific~\cite{Birge}. The most generic and common for
all the conductors mechanism stems from the electron-electron
interaction~\cite{AAK}. It yields $1/\tau_\phi\sim
T(G_Q/\sigma_0)\ln(G_Q/\sigma_0)$ and provides the temperature
dependence of the conductivity,
\begin{equation}
\sigma=\sigma_0-(G_Q/2\pi)\ln (T^*/T) \label{eq:1}
\end{equation}
with $T^*\sim\sigma_0/G_Q\tau$. The typical area under an electron
trajectory which barely preserves coherence, $L_\phi^2 =
D\tau_\phi$, depends on the electron diffusion constant. Magnetic
field $B$ significantly affects the WL correction if the
corresponding magnetic flux through a contour of area $L_\phi^2$
exceeds the quantum $\Phi_0$. This makes the magnetoresistance
measurement a useful tool for the investigation of the electron
interference.

To model a granular medium, we consider a regular two-dimensional
array of grains of size $d$ connected by tunnel junctions. The
grains have internal disorder, but are characterized by
conductance far exceeding the conductance $g_{\rm T}G_Q$ of a
single tunnel junction. The classical conductivity of a square
array is thus $\sigma_0=G_Qg_{\rm T}$.  It
corresponds~\cite{Beloborodov01} to the effective electron
diffusion constant $D = \pi^{-1} g_{\rm T}\delta d^2$.  In the
absence of phase relaxation, an electron may pass through any
number of junctions coherently. It will result in a diverging WL
correction, just like in a homogeneous conductor. The
electron-electron interaction limits the phase relaxation time,
yielding $1/\tau_\phi\sim T/g_{\rm T}$. As long as the
corresponding length $L_\phi\sim d\sqrt{g^2_{\rm T}\delta/T}$
significantly exceeds $d$, an electron may return to a grain
coherently after passing many junctions, and the inhomogeneity of
the granular medium is irrelevant. The comparison of $L_\phi$ with
$d$ defines a crossover temperature,
\begin{equation}
T_{\rm cr}=g^2_{\rm T}\delta. \label{eq:2}
\end{equation}
Roughly, above the crossover temperature the electron trajectories
contributing to the WL do not cross
more than a single junction.
In this regime granular medium behaves similarly to a single grain
connected to highly conducting leads by tunnel junctions of
conductance $g_{\rm T}$.

The WL correction
at $T\gtrsim T_{\rm cr}$ comes from electron trajectories that pass
%once
%twice
through a single tunnel junction.
% (leaving and re-entering a grain).
Electrons moving along longer trajectories, which include more
junctions, have much smaller probability of a phase-coherent return.
We find that already the shortest
inter-grain trajectories (see Fig.~2 in Section \ref{conductivity}) provide the temperature dependence of the WL
correction,
\begin{equation}
\delta\sigma_{\rm WL}=-A
\frac{T_{\rm
cr}}{T}, \label{eq:3}
\end{equation}
with $A$ being a geometry-dependent coefficient of order one.
In deriving Eq.~(\ref{eq:3}), we assume that $g_{\rm T}$ is much
smaller than the number of channels
in the inter-grain tunnel junction, although $g_{\rm T}\gg 1$.

Equation (\ref{eq:3}) does not account for the phase relaxation rate
within the grains. At a sufficiently high temperature $T\gtrsim T^*$
the latter exceeds the electron escape rate $g_{\rm T}\delta$ from a
grain, which leads to a suppression of the WL correction below the
value Eq.~(\ref{eq:3}). The characteristic scale $T^*$ here depends on
the intra-grain phase relaxation mechanism. Assuming that it is due to
the electron-electron interaction~\cite{Sivan}, and that the
dimensionless conductance of the grain $g_{\rm gr}$ is large, $g_{\rm
  gr}\gtrsim g_{\rm T}^2$, we find
\begin{equation}
T^*\sim T_{\rm cr}\frac{g_{\rm gr}}{g_{\rm T}^2}\sqrt{g_{\rm T}}.
\label{t-star}
\end{equation}
In a more exotic case of a smaller grain conductance, $g_{\rm T}^2\gg
g_{\rm gr}\gg g_{\rm T}$, the temperature $T^*$ still exceeds
significantly $T_{\rm cr}$, but the specific relation between the two
temperature scales depends on the grain shape, and is different for
disk-like or dome-like grains.

We turn now to the discussion of the magnetic field effect on the weak
localization in the granular medium. To determine the characteristic
field suppressing the interference correction to conductivity, we need
to estimate the directed area covered by a typical closed electron
trajectory~\cite{ABG}. For a single grain, such area is of the order
$d^2\sqrt{g_{\rm gr}/g_T}$ and is limited by the electron dwelling
time. At low temperatures, $T\ll T_{\rm cr}$, the number of grains
visited by an electron having a typical closed trajectory, is of the
order of $T_{\rm cr}/T$. The single-grain directed areas have random
signs, so the estimate for the full directed area is
\begin{equation}
S_{\rm eff}\sim
d^2\sqrt{\frac{T_{\rm cr}g_{\rm gr}}{T g_{\rm T}}}+d^2\frac{T_{\rm
cr}}{T}\,.
\label{area}
\end{equation}
The first term here corresponds to the sum of the directed areas
under the electron trajectory within the grains visited by
electron; the second, conventional~\cite{Gershenson}, term comes
from the fact that the ``visited'' grains form a closed contour of
an area $L_\phi^2$, see Fig.~1. The characteristic level of the
field necessary to affect $\delta\sigma_{\rm WL}$ is found from
the condition $S_{\rm eff}B_\phi\sim\Phi_0$. We see now, that even
within the temperature range $T\lesssim T_{\rm cr}$, the
granularity of the material matters.
\begin{figure}
\includegraphics[width=2.5in]{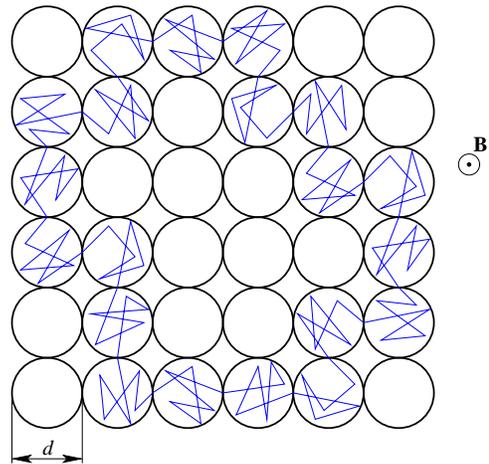}
\caption{A typical diffusive trajectory in a granular array. The
  directed  area $S_{\rm eff}$ consists of two components. The first
  one is the combined contribution of separate grains, see the
  first term in Eq.~(\ref{area}). The second component is the area under
  the coarse-grained trajectory, which is the counterpart of the directed
  area under an electron trajectory in a homogeneous disordered sample.}
\end{figure}

At the lowest temperatures the characteristic field coincides with that
of a film with the corresponding value of diffusion coefficient,
\begin{equation}
B_\phi\sim\frac{\Phi_0}{d^2}\frac{T}{T_{\rm cr}},\quad
T\ll T_{\rm cr}\frac{g_{T}}{g_{\rm gr}}.
\label{field1}
\end{equation}
At higher temperatures, the characteristic field is
\begin{equation}
B_\phi\sim \frac{\Phi_0}{d^2}\sqrt{\frac{Tg_{\rm T}}{T_{\rm cr}g_{\rm
      gr}}}\;,\quad
T_{\rm cr}\frac{g_{T}}{g_{\rm gr}}\ll T\ll T_{\rm cr}.
\label{eq:4}
\end{equation}
The higher the applied field, the shorter are the trajectories
contributing to the interference correction, and the smaller the
correction is. Such trajectories span only a single grain provided the
field $B$ is of the order or higher than
\begin{equation}
B_\phi^{\rm sg}=\frac{\Phi_0}{d^2}\sqrt{\frac{g_{\rm T}}{g_{\rm gr}}}.
\label{eq:5}
\end{equation}
At $B\gg B_\phi^{\rm sg}$, even the single-grain Cooperon is
suppressed. Consequently, Eq.~(\ref{eq:5}) defines the characteristic
field suppressing at $T\gtrsim T_{\rm cr}$ the WL correction
Eq.~(\ref{eq:3}), which stems from the transitions within the closest
grain pairs.

To develop a quantitative theory of the interference correction,
we derive the expression for the weak localization correction and
adapt the Cooperon equation for granular medium.

\section{Conductance and weak localization in a granular array}

Tunnel junctions between metallic grains are described adequately
by a model with an infinitely large number of weakly-conducting
channels.
Within this model, one can use the tunneling Hamiltonian formalism
for evaluation of the conductivity of the granular array. In this
formalism, tunneling between the grains $m$ and $n$ is described
by the Hamiltonian
\begin{eqnarray} \label{tunham}
& & \hat{H}_{\rm T} = \sum_{\alpha\beta\sigma} t_{\alpha\beta}
e^{iV_{mn}t}
\hat a_{\alpha\sigma}^{\dagger} \hat a_{\beta\sigma} + h. c. =
\sum_{\sigma} \int_m d\bm{r}_1 \\
& & \times \int_n d\bm{r}_2 t (\bm{r}_1,
\bm{r}_2) e^{iV_{mn}t} \hat \psi_{m\sigma}^{\dagger} (\bm{r}_1) \hat
\psi_{n\sigma} (\bm{r}_2) + h.c. \ , \nonumber
\end{eqnarray}
where the points $\bm{r}_1$ and $\bm{r}_2$ belong to the grains $m$
and $n$, respectively, $V_{mn}$ is the voltage applied to the barrier,
$\alpha \in m$ and $\beta \in n$ are exact single-particle states,
and $\sigma$ is the spin index. In the limit of thin barrier, the
tunnel amplitude $t$ significantly deviates from zero only if the
vectors $\bm{r}_1$ and $\bm{r}_2$ refer to two closest to each other
points at opposite sides of the interface,
\begin{equation} \label{tunamp2}
t (\bm{r}_1, \bm{r}_2) = a \delta (y - y') \partial_z
\partial_{z'} \delta (z) \delta(z').
\end{equation}
Here the coordinate $y$ runs along the interface $S$, and transverse
coordinates $z$ in the grain $m$ and $z'$ in the grain $n$ are defined
in such a way that at the interface $z = z' = 0$. (We wrote Eq.
(\ref{tunamp2}) for the planar geometry, generalization to
three-dimensional arrays is straightforward). The constant $a$ is of
the order of magnitude of $\sqrt{{\cal T}}/\nu k_F$, where $\nu$ is
the electron density of states of the material of the
grains, and ${\cal T}$ is the transmission
coefficient of the barrier. The
numerical factor in $a$ can be related to the measurable quantity,
the barrier conductance $g_{\rm T}$.
Using Eq.~(\ref{tunamp2}), one may express the tunnel amplitude in
terms of the eigenfunctions $\chi_{\alpha}$ and $\chi_{\beta}$ of an
electron in the grains $m$ and $n$, respectively (see, {\em e.g.},
Ref.~\onlinecite{Houzet}),
\begin{equation} \label{tunamp1}
t_{\alpha\beta} =
a\int_S dy
\partial_z \chi^*_{\alpha} (y,z) \partial_{z'} \chi_{\beta} (y,z')
\vert_{z = z' = 0}\ .
\end{equation}

The current through the contact is defined as
$\hat I = -e \hat{\dot{N}}_m = -ie [ \hat{H}_{\rm T} , \hat{N}_m ]$, where
$\hat{N}_m$ is the number of particles in the grain $m$,
\begin{displaymath}
\hat{N}_m = \sum_{\alpha\sigma} \hat{a}^{\dagger}_{\alpha\sigma}
\hat{a}_{\alpha\sigma} \ .
\end{displaymath}
Calculating the average current through the barrier, we obtain
\begin{eqnarray*}
I (t) = -e \ \mbox{Re} \int d\bm{r}_1 d\bm{r}_2 \left\{
t(\bm{r}_1, \bm{r}_2) G^K_{nm} (\bm{r}_2, \bm{r}_1, t, t) e^{iV_{mn}t}
\right.
\\
\left. - t^*(\bm{r}_1, \bm{r}_2) G^K_{mn} (\bm{r}_2, \bm{r}_1, t, t)
e^{-iV_{mn}t} \right\} \ ,
\end{eqnarray*}
where $G^{K}$ is the Keldysh component of the matrix Green's function,
and the subscripts $m$ and $n$ are introduced for convenience, in
order to indicate which grain points $\bm{r}_1$ and $\bm{r}_2$ belong
to. We need now to calculate the function $G^K$ by perturbation theory
in tunneling Hamiltonian. Let us first discuss the first order and
calculate the average conductance. Using the standard technique
\cite{Rammer}, we obtain for the current in the frequency
representation (terms which do not depend on the time difference would
correspond to the Josephson effect and thus are dropped)
\begin{eqnarray} \label{currbas1}
I (\omega) & = & 2e \ \mbox{Re} \int d\bm{r}_1 \dots d\bm{r}_4
\frac{d\omega}{2\pi} t^*(\bm{r}_1, \bm{r}_2) t(\bm{r}_3, \bm{r}_4) \\
& \times & \mbox{Tr} \left[ \hat \tau_x \hat G_m (\bm{r}_1, \bm{r}_3,
\omega + eV_{mn}) \hat G_n (\bm{r}_4, \bm{r}_2, \omega) \right],
\nonumber
\end{eqnarray}
where $\tau_x$ is the Pauli matrix in the Keldysh space, $G_{n} \equiv
G_{nn}$, and we use the standard representation
\begin{eqnarray*}
\hat G = \left( \begin{array}{cc} G^R & G^K \\ 0 & G^A \end{array}
\right) \ .
\end{eqnarray*}
In the linear regime, it suffices to use the equilibrium function
here, $G^K (E) = \tanh (E/2T) (G^R (E) - G^A (E))$.  Expressing the
Green's functions in terms of the exact eigenfunctions,
calculating the energy integrals, and substituting the transmission
amplitudes (\ref{tunamp2}), we obtain $I = \sigma V_{mn}$ for the
inter-grain current, and $\sigma = (e^2/\pi)g_{\rm T}$ for the Drude
conductivity of the granular array. The introduced here dimensionless
inter-grain tunneling conductance is
\begin{eqnarray} \label{tuncond}
& & g_{\rm T} = 4\pi^2 \vert a \vert^2 \int_S dy dy' \\
& & \left\vert \left\langle \sum_{\alpha}
\partial_z \chi_{\alpha} (yz)
\partial_{z'} \chi^*_{\alpha} (y'z') \vert_{z = z' = 0}
\delta(\xi_{\alpha}) \right\rangle \right\vert^2 \nonumber \\
& & = 4\vert a \vert^2 \int_S dy dy' \left[ \partial_z \partial_{z'}
\mbox{\rm Im} \ \left\langle G^R (yz;y'z') \right\rangle \vert_{z = z'
= 0} \right]^2 \ , \nonumber
\end{eqnarray}
where $\xi_\alpha$ are the exact energy eigenvalues measured from the
Fermi level in a grain, and angular brackets mean impurity averaging
within a grain (the eigenfunctions in different grains are not
correlated). In the
last equation, $\langle G^R \rangle$ is the impurity-averaged Green's
function evaluated at the Fermi energy. It is represented as the
density of states $\nu$ multiplied with a dimensionless function
rapidly decaying with the distance $y - y'$. The characteristic
length of that decay is given by the Fermi wavelength, and the
integral in Eq.~(\ref{tuncond}) is converging rapidly. Therefore the
dimensionless function of $y - y'$ may be evaluated within the
free-electron approximation~\cite{Houzet}. The precise shape of this
function is not important for our purposes. Equation (\ref{tuncond})
thus relates the tunnel conductance to the previously introduced
constant $a$.

\begin{figure} \label{cooperon}
\includegraphics[height=2.5in]{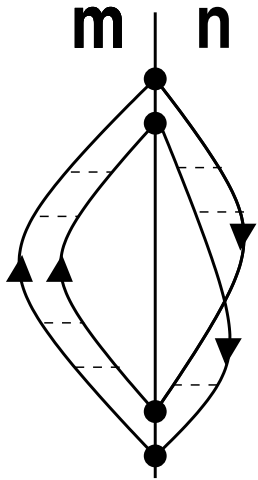}
\caption{Second-order correction to the conductivity. Black circles
represent the tunnel amplitudes, and dashed lines denote impurity
scattering inside the grains}
\end{figure}

We proceed now with the evaluation of weak localization correction.
The next-order contribution to the current (Fig.~2)
contains four tunnel amplitudes and four Green's functions with the
Keldysh structure $\mbox{Tr} \ (\hat\tau_x \hat G \hat G \hat G \hat
G)$, where $\hat G$ is the matrix Green's function in the Keldysh
space. The trace of a product of several Green's function can only
have the following structure: several
first functions are retarded, followed by one Keldysh function and
then a number of advanced functions. Thus, we have the combination of
the type $G^K G^A G^A G^A + G^R G^K G^A G^A + G^R G^R G^K G^A + G^R
G^R G^R G^K$. However, the second and third terms in this
combination are considerably greater than the other two, since the
impurity scattering inside the grains is the most effective if the
impurity line connect advance and retarded, advanced and Keldysh, and
Keldysh and retarded Green's functions, but not two retarded or two
advanced ones. Thus, retaining only these two
terms\cite{AAK,AA}, we
express the weak localization correction in terms of the Cooperon
$C_{mn}$ in the time representation,
\begin{eqnarray} \label{wlstart}
& & \delta \sigma_{\rm WL} = -\frac{2e^2}{\pi} \mbox{Re}\ \int
d\bm{r}_1 \dots d\bm{r}_4 t^*(\bm{r}_1, \bm{r}_2) t(\bm{r}_3,
\bm{r}_4) \nonumber \\
& & \times \int_{-\infty}^{\infty} dt C_{mn} (\bm{r}_1, \bm{r}_4;
\bm{r}_3, \bm{r}_2; t, -t) \ ,
\end{eqnarray}
where the subscripts of the Cooperon indicate that it starts and ends
in the grains $m$ and $n$, respectively.
%According to Eq.~(\ref{tunamp2}), points $\bm{r}_1$ and $\bm{r}_2$ are
%separated by an infinitely thin interface while belonging to grains
%$m$ and $n$, respectively; the same is true for points $\bm{r}_4$ and
%$\bm{r}_3$. Therefore
Note that due to the structure of the tunneling amplitudes
$t(\bm{r},\bm{r}^\prime)$, point $\bm{r}_1$ is just across the
barrier from point $\bm{r}_2$ and similarly point $\bm{r}_3$ is
across the barrier from point $\bm{r}_4$.
The Cooperon $C$ can be presented in the form
%effectively depends only on two coordinates,
\begin{eqnarray*}
& & C_{mn} (\bm{r}_1, \bm{r}_4; \bm{r}_2, \bm{r}_3) =
\frac{\pi}{\nu} \mbox{Im}\ \left\langle G^R_m (\bm{r}_1 - \bm{r}_4)
\right\rangle \\
& & \times  \mbox{Im} \left\langle G^R_n (\bm{r}_2 - \bm{r}_3)
\right\rangle \tilde C_{mn} (\bm{r}_1, \bm{r}_3) \ .
\end{eqnarray*}
Rapid decay of functions $\langle G^R \rangle$ with the distance
between the corresponding arguments makes points in pairs $\bm{r}_1$,
$\bm{r}_4$ and $\bm{r}_2$, $\bm{r}_3$ in the spatial integral of
Eq.~(\ref{wlstart}) to be within the Fermi wavelength from each other.
On the other hand, the Cooperon $\tilde C (\bm{r}_1, \bm{r}_3)$ is
generally a long-range function. Provided we are interested in times
long compared to the intra-grain diffusion time, $\tilde C$ almost
does not change while
%$\bm{r}$ or $\bm{r}'$
its coordinates vary within respective grains.  However, $\tilde
C_{mn} (\bm{r}_1, \bm{r}_3)$ with $m\neq n$ may differ significanlty
from the value of single-grain Cooperon ($m=n$). Substituting this
coarse-grained Cooperon $\tilde C_{mn}$ into Eq. (\ref{wlstart}) and
taking into account Eqs.  (\ref{tunamp2}) and (\ref{tuncond}), we
obtain
\begin{equation} \label{WLbase1}
\delta\sigma_{\rm WL} = \frac{e^2g_{\rm T}}{2\pi\nu^2} \mbox{Re}
\int_{-\infty}^{\infty} dt \tilde C_{mn}(t, -t) \
\end{equation}
with $m$ and $n$ being the neighboring grains.
Note that Eq. (\ref{WLbase1}) is valid for any dimension, not just in
2D.

The form (\ref{WLbase1}) of weak localization
correction is valid provided the phase coherence between the
grains barely survives, and $\tilde C_{mn}\ll \tilde C_{nn}$ at
$m\neq n$. This limit is realized at a sufficiently high
temperature, $T\gg T_{\rm cr}$. Note also that the performed
derivation, unlike the consideration of, {\it e.g.},
Ref.~\onlinecite{VA} assumes the limit of large number of channels taken
at fixed value of $g_{\rm T}$.

To consider phase relaxation in a granular array, we derive now the
proper equation for $\tilde C_{nm}$ in a granular medium.

%{\bf then your abbreviated write-up: (1) use of $dN/dt$ to define
%current; expansion to the first order in tunneling to get the
%conductance; (2) next two orders of expansion in order to get the
%WL correction in terms of $C_{mn}$. Figure - diagram for the WL
%correction}

\section{Cooperon in a granular array}

Cooperon describes the probability amplitude of electron return
and in the case of a homogeneous medium with electron diffusion
coefficient $D$ obeys the equation
\begin{widetext}
\begin{eqnarray} \label{Cooperon0}
\left\{ \frac{\partial}{\partial t} - D \left[
\frac{\partial}{\partial \bm{r}} - i\frac{e}{c} \bm{A}
(\bm{r}, -t/2) - i\frac{e}{c} \bm{A} (\bm{r}, t/2)\right]^2
\right\} \tilde{C} (\bm{r}, \bm{r}'; t, t') = \delta
\left(\bm{r} - \bm{r}' \right) \delta (t-t').
\end{eqnarray}
\end{widetext}
Here the vector potential $\bm{A}$ accounts for the fluctuating
electric fields representing the effect of electron-electron
interactions, and should be considered as a Gaussian classical random
variable with zero average.

In order to adapt Eq.~(\ref{Cooperon0}) to
the case of a granular medium, it is convenient to perform a gauge
transformation, after which the fluctuating fields are described by a
random scalar potential $\varphi (\bm r, t)$, rather than by the
vector potential $\bm{A} (\bm{r}, t)$,
\begin{displaymath}
\bm{A} (\bm{r}, t) = c \int^t \nabla_{\bm{r}} \varphi (\bm
{r}, t') dt'
\end{displaymath}
(we assume there are no magnetic fields applied to the system).
Defining the Cooperon $C (\bm{r}, \bm{r}'; t, t'))$ in the new gauge
by the relation
\begin{eqnarray} \label{gauge}
& & \tilde{C} (\bm{r}, \bm{r}'; t, t') =
C (\bm{r}, \bm{r}'; t, t')) \\
& & \times \exp \left\{ ie \int^{t/2} \varphi(\bm{r}, t'') dt'' - ie
\int^{t'/2} \varphi(\bm{r}', t'') dt'' \right. \nonumber \\
& & \left. + ie \int^{-t/2}
\varphi(\bm{r}, t'') dt'' - ie \int^{-t'/2} \varphi(\bm{r}', t'')
dt''\right\}\ \ \nonumber ,
\end{eqnarray}
we obtain the equation
\begin{widetext}
\begin{eqnarray} \label{Cooperon00}
\left\{ \frac{\partial}{\partial t} + \frac{ie}{2}
\varphi(\bm{r}, t/2) - \frac{ie}{2} \varphi(\bm{r}, -t/2) - D
\frac{\partial^2}{\partial \bm{r}^2} \right\} C (\bm{r},
\bm{r}'; t, t') = \delta \left(\bm{r} - \bm{r}' \right)
\delta (t-t').
\end{eqnarray}
\end{widetext}
Note that $\tilde{C} (\bm{r}, \bm{r}; t, -t) = C (\bm{r}, \bm{r};
t, -t)$, and thus $C$ can be used instead of $\tilde{C}$ in for
evaluation of the WL correction (\ref{WLbase1}).

Returning to the consideration of a granular array, we assume that the
intra-grain conductance is high, $g_{\rm gr}\gg g_{\rm T}$. Then the
fluctuating potential $\varphi(\bm{r}, t)$ does not vary from point to
point within a single grain, while exhibiting random fluctuations of
the inter-grain potential differences. This allows us to coarse-grain
function $\varphi(\bm{r}, t)$, replacing its dependence on ${\bm r}$
by the dependence on the grain number $n$. We also can simplify the
spatial dependence of the Cooperon $C (\bm{r}, \bm{r}'; t, t'))$, in
case we are interested in times long compared to the
intra-grain diffusion time. Indeed, in that case $C$ does not change
while $\bm{r}$ or $\bm{r}'$ vary within a grain. Therefore, the
dependence of the Cooperon on the coordinates may be replaced by the
dependence on the grain numbers $n$ and $n'$ which the coordinates
$\bm{r}$ and $\bm{r}'$ belong to. The resulting coarse-grained
equation for the Cooperon reads:
\begin{eqnarray} \label{Cooperon1}
& & \left\{ \frac{\partial}{\partial t} + \frac{ie}{2} \varphi_n(t/2)
- \frac{ie}{2} \varphi_n (-t/2) +
N\frac{g_{\rm T}\delta}{\pi} \right\} C_{nn'}
(t,t') \nonumber \\
& & - \frac{g_{\rm T}\delta}{\pi} \sum_k C_{kn'} (t,t') =
\delta_{nn'}\delta(t-t').
\end{eqnarray}
Here $N$ is the number of junctions to a single grain ({\it i.e.},
the coordination number of the lattice of grains; $N=4$ for a
two-dimensional square lattice), and the summation in the last
term on the left-hand side runs over $N$ nearest neighbors $k$ of
the grain $n$.

Equations (\ref{WLbase1}) and (\ref{Cooperon1}) provide a
convenient starting point for evaluation of the weak localization
correction at temperatures $T\ll T^*$, see Eq.~(\ref{t-star}). At
higher temperatures, the spatial dispersion of the fluctuating
potentials and of the Cooperon inside a grain becomes important.

The temperature domain $T\lesssim T^*$ is separated in two
characteristic regions by the scale $T_{\rm cr}$, Eq.~(\ref{eq:2}). At
$T\ll T_{\rm cr}$, the dependence of Cooperon $C$ on $n-n'$ is smooth,
and the finite difference equation (\ref{Cooperon1}) can be replaced by
the corresponding differential equation, which essentially returns one
to the continuous-medium case, see Eq.~(\ref{Cooperon00}). Weak
localization corrections in this case are studied in detail in Refs.
\onlinecite{AAK,AA}. Below we
concentrate on the temperatures above the crossover.

\section{Quantum correction to conductivity above the crossover
temperature} \label{conductivity}

In the temperature regime of interest,
\begin{equation}
T^*\gg T\gtrsim T_{\rm cr},
\label{t-interval}
\end{equation}
as we have explained in the Introduction, electron trajectories are
classified according to the number of tunnel junctions they cross ---
the longer are the trajectories, the less significant are their
contributions. It means that the matrix $C_{nn'}$ rapidly decays away
from the diagonal. The biggest matrix elements are $C_{nn}$, and the
most important trajectories are those which do not leave the
grain. Eq. (\ref{WLbase1}) implies that these trajectories do not
contribute to the weak localization correction, and one needs to
consider the next-order contribution coming from trajectories crossing
a single junction once. This leads us to Eq.~(\ref{eq:3}) and also
allows us to verify the existence of the crossover temperature
Eq.~(\ref{eq:2}).

At $T\gg T_{\rm cr}$ we expect strong fluctuations of the potential
differences between the grains, making coherent returns of an
electron to the grain of its origin improbable. The returns are
described by the term in Eq.~(\ref{Cooperon1}) containing the sum over
$k$. Neglecting that term, we find for the diagonal component
$C_{nn}(tt')$ of the Cooperon
\begin{eqnarray}
\label{zeroorderCoop}
& & C_{nn}^{(0)} (t,t') = \theta(t-t') \exp
\left\{ - N\frac{g_{\rm T}\delta}{\pi} (t-t') \right. \nonumber \\
& & - \left. ie \int_{t'/2}^{t/2}
\varphi_{n}(t'') dt'' - ie \int_{-t'/2}^{-t/2} \varphi_{n}(t'') dt''
\right\} .
\end{eqnarray}
The phase factors here reflect the specific gauge we used in
Eq.~(\ref{Cooperon1}).

Next, we write the Cooperon equation (\ref{Cooperon1}) for the
nearest-neighbor sites $m$ and $n$,
\begin{eqnarray}
\label{Cooperon11} & & \left\{ \frac{\partial}{\partial t} +
\frac{ie}{2} \varphi_m(t/2) - \frac{ie}{2} \varphi_m (-t/2) +
N\frac{g_{\rm T}\delta}{\pi} \right\} C_{mn}
(t,t') \nonumber \\
& & = \frac{g_{\rm T}\delta}{\pi} C^{(0)}_{nn} (t,t').
\end{eqnarray}
The terms with $C_{n', n}$ describing the grains $n'$ separated from
$n$ by two tunnel junctions, are small and can be omitted in this
approximation. Using Eqs. (\ref{zeroorderCoop}) and
(\ref{Cooperon11}), we obtain for the neighboring grains $m$ and
$n$
\begin{eqnarray}
\label{offdiag} & & C_{mn} (t,t') = \frac{g_{\rm T}\delta}{\pi}
\theta(t-t') \exp \left\{ - N\frac{g_{\rm T}\delta}{\pi} (t-t')
\right\} \int_{t'}^{t} dt_1 \nonumber
\\
& & \times \exp \left\{ - ie \int_{t'/2}^{t_1/2} \varphi_{n}(t'')
dt'' - ie
\int_{-t'/2}^{-t_1/2} \varphi_{n}(t'') dt'' \right. \nonumber \\
& & \left. + ie \int_{t/2}^{t_1/2} \varphi_{m}(t'') dt'' + ie
\int_{-t/2}^{-t_1/2} \varphi_{m}(t'') dt'' \right\} \ .
\end{eqnarray}
This expression has to be averaged over the Gaussian fluctuations
of the field $\varphi$. Its correlation function is defined by the
fluctuation-dissipation theorem and reads~\cite{AB02}
\begin{displaymath}
e^2 \langle \varphi \varphi \rangle (\bm{q}, \omega) = - \mbox{\rm Im}
\ \frac{2T}{\omega} \frac{1}{\Pi(\bm{q}, \omega)},
\end{displaymath}
where $\Pi$ is the polarization operator, calculated for a granular
medium in Ref. \onlinecite{Beloborodov01}. In the space-time
representation, we obtain
\begin{eqnarray}
\label{phicorr}
& & e^2 \left\langle \varphi_n (t) \varphi_{n'} (t') \right\rangle =
 \frac{\pi Td^2}{g_{\rm T}} \delta(t-t') \nonumber \\
& \times & \int \frac{d\bm{q}}{(2\pi)^d}
 \frac{e^{i\bm{q} (\bm{n}-\bm{n'})}}{\sum_a (1 - \cos q_a d)},
\end{eqnarray}
where the summation in the denominator is carried over all available
Cartesian components $a = x,y$, and ${\bm q}_a$ are the basis vectors of
square lattice of grains.

Performing the averaging in Eq.~(\ref{offdiag}) with the help of
Eq.~(\ref{phicorr}) is cumbersome but straightforward, since for
Gaussian fields $\langle \exp(i\dots) \rangle =
\exp(-\langle\dots\rangle/2)$. For the weak localization
correction, we obtain
\begin{eqnarray}
& & \frac{\delta\sigma_{\rm WL}}{\sigma_0} = -A
\frac{g_{\rm T}\delta}{T}, \label{firstordWL} \\
& & A \equiv \frac{1}{N^2V} \sum_a
\left( \int\frac{d\bm{q}}{(2\pi)^d}
 \frac{1 - \cos q_a d}{\sum_a (1 - \cos q_a d)} \right)^{-1}\,,\nonumber
\end{eqnarray}
where $V$ is the volume of the grain ($V = d^2$ in 2D). Note that
the interference correction Eq.~(\ref{firstordWL}) depends on
temperature and on the type of lattice the grains form.
%, in contrast to the single-grain contribution (\ref{zeroordWL}).
The dependence on the lattice type comes through the coefficient
$A$; for a square 2D lattice we find $A=1/4$. One can easily
generalize the evaluation of $A$ onto the case of a triangular and
more complicated lattices, eventually even describing disordered
media like ceramics.

\section{Magnetic field effect}

As the estimate Eq.~(\ref{area}) suggests, the action of the magnetic
field on Cooperon is two-fold. A part of the Cooperon suppression
comes from the intra-grain electron motion, and another part stems
from the magnetic field effect on the inter-grain coherence.  Since
$g_{\rm gr}\gg g_{\rm T}$, the interesting range of the fields
corresponds to a small flux penetrating a grain, $Bd^2\ll\Phi_0$.  We
can then consider the effect of magnetic field on the Cooperon within
a grain perturbatively. To implement the perturbation theory, it is
convenient to use a ``tailored'' to the grains shape gauge of the
magnetic field $B$. For definiteness, we concentrate on the case of a
two-dimensional array of ``flat" grains connected by point-like
tunneling contacts, see Fig.~1. We define the gauge for the points
within the grains by the relations
\begin{equation}
{\bm A}_\phi({\bf r})={\bm\tau}\times\nabla\psi_n({\bf r})+{\bm
  A}_n\,,\;\nabla^2\psi_n=B\,,\; \psi_n ({\bm r}\in b_n)=0.
\label{gauge1}
\end{equation}
Here ${\bm \tau}$ is the normal to the plane of the grains,
and $b_n$ is the
boundary of the $n$--th grain. The second and third relations in
Eq.~(\ref{gauge1}) fully define the boundary problem for a scalar
function $\psi_n({\bm r})$.  The constants ${\bm A}_n$ are tuned in
such way that the vector-potential is continuous at the points of
contact between the grains. Up to a discrete analogue of the gradient
of a scalar function, these constants are determined fully by the
solution of the boundary problems for all $\psi_m({\bm r})$. It is
clear that the discrete version of curl applied to ${\bm A}_n$ must be
equal ${\bm B}$ upon averaging over the array; the characteristic
difference $A_n-A_{m}$ for two nearby junctions is of the order
$A_n-A_{m}\sim B\cdot d$.

In the definition of the Cooperon, it is convenient to present again
the coordinates as pairs $\{{\bm r},n\}$ and $\{{\bm r}',n'\}$ which
point explicitly to the label of grains the two points ${\bm r}, {\bm
r}'$ belong to. In addition, we multiply the Cooperon defined in
Eq.~(\ref{gauge}) by yet one more
%an additional
gauge factor,
\begin{equation}
C_{mn} (\bm{r}, \bm{r}'; t, t')=C^\phi_{mn} (\bm{r} , \bm{r}'; t,
t')\exp(i{\bm A}_m\cdot{\bm r}-i{\bm A}_{n}\cdot{\bm r}').
\label{gauge2}
\end{equation}
In these new notations, the equation for Cooperon in the absence of
tunneling has the form
\begin{widetext}
\begin{eqnarray} \label{Cooperon10}
\!\left\{\! \frac{\partial}{\partial t} + \frac{ie}{2}
\varphi(\bm{r}, t/2) - \frac{ie}{2} \varphi(\bm{r}, -t/2) - D_{\rm gr}
\left(\frac{\partial}{\partial \bm{r}}-i\frac{e}{c}{\bm \tau}\times
\nabla\psi_m({\bm r})\right)^2\! \right\} C^\phi_{mn} (\bm{r},
\bm{r}'; t, t') = \delta_{mn}\delta \left(\bm{r} - \bm{r}' \right)
\delta (t-t'),
\end{eqnarray}
\end{widetext}
where $D_{\rm gr}\sim d^2g_{\rm gr}\delta$ is the diffusion coefficient
within the grain (here $g_{\rm gr}\delta$ is the Thouless energy for
the electron motion within a grain). With the defined gauge
Eq.~(\ref{gauge1}), the normal to the boundary component of ${\bm
  A}({\bm r})$ is zero.  Thus, the magnetic field does not affect the
boundary conditions for Cooperon, {\it i.e.} the normal component of
$\partial C^\phi/\partial{\bm r}$ at the boundary is zero.

As long as the flux piercing one grain is small compared with the unit
quantum, $Bd^2\ll \Phi_0$, we may treat the effect of a magnetic field
within a grain perturbatively. Considering the low-energy limit, $T\ll
g_{\rm gr}\delta$, and taking into account the boundary conditions for
$C^\phi$, we start perturbations from ${\bm r}$--independent Cooperon
$C^\phi_{mn} (t, t')$. In the presence of inter-grain tunneling, the
corresponding generalization of Eq.~(\ref{Cooperon1}) reads
\begin{widetext}
\begin{eqnarray} \label{Cooperon12}
& & \left\{ \frac{\partial}{\partial t}
+\alpha g_{\rm gr}\delta\left(\frac{Bd^2}{\Phi_0}\right)^2
+\frac{ie}{2} \varphi_m(t/2) - \frac{ie}{2} \varphi_m (-t/2)
+N\frac{g_{\rm T}\delta}{\pi} \right\} C^\phi_{mn} (t,t') \nonumber \\
& & - \frac{g_{\rm T}\delta}{\pi} \sum_k
e^{i{\bm r}_{km}\cdot({\bm A}_k-{\bm A}_m)}C^\phi_{kn}(t,t') =
\delta_{mn}\delta(t-t').
\end{eqnarray}
\end{widetext}
Here the magnetic field dependence
\[
\alpha g_{\rm gr}\delta\left(\frac{Bd^2}{\Phi_0}\right)^2=
\frac{D_{\rm gr}}{d^2} \frac{e^2}{c^2}
\int_{\rm grain} d^2{\bm r} |\nabla\psi_n({\bm r})|^2
\]
comes from the $\psi_n$--dependent term in Eq.~(\ref{Cooperon10})
integrated over the volume of a single grain; $\alpha\sim 1$ is the
dimensionless coefficient depending on the grains shapes. The vector
${\bm r}_{kn}$ points to the junction between grains $k$ and $n$.

The discreteness of the medium is adequately accounted for by the
structure of Eq.~(\ref{Cooperon12}). However, the discreteness is not
important in the domain of low temperatures and relatively low fields,
$T\ll T_{\rm cr}$ and $B\ll B_\phi^{\rm sg}$. There we can replace the
left-hand side of Eq.~(\ref{Cooperon12}) by its gradient
expansion. After the expansion and replacement of the grain number $n$
by the corresponding coarse-grained coordinate ${\bm R}$, we find
\begin{eqnarray}
&& \left\{ \frac{\partial}{\partial t}
+\alpha g_{\rm gr}\delta\left(\frac{Bd^2}{\Phi_0}\right)^2
-D\left[\nabla_{\bm R}-\frac{ie}{c}{\bm A}({\bm R})\right]^2
\right. \label{Cooperon13}\\
&&\left.+\frac{ie}{2} \varphi({\bm R}, t/2) - \frac{ie}{2}
\varphi ({\bm R}, -t/2)
\right\} C^\phi({\bm R},{\bm R}';t,t') \nonumber\\
&& = \delta ({\bm R}-{\bm R}')\delta(t-t').\nonumber
\end{eqnarray}
The second term in the left-hand side here reflects the suppression of
interference by the magnetic flux penetrating the grains.
Apart from that term and from the value of the effective diffusion
constant $D=\pi^{-1}g_{\rm T}\delta d^2$, which reflects the
granularity of the medium, this equation is identical to that of a
homogeneous thin film. Using the known results~\cite{Gershenson} for
the films, we find the magnetoresistance of a granular array,
\begin{eqnarray}
\label{MR}
&&\delta\sigma_{\rm MR}(B, T)=\delta\sigma_{WL}(B,T)-
\delta\sigma_{WL}(0,T)\\
&&=\frac{e^2}{2\hbar}\left\{
\ln\left[\frac{T}{T_{\rm cr}}
+\frac{g_{\rm gr}}{g_{\rm T}}\left(\frac{Bd^2}{\Phi_0}\right)^2
+\frac{Bd^2}{\Phi_0}\right]-\ln\frac{T}{T_{\rm cr}}\right\}\nonumber
\end{eqnarray}
(we dispensed with the factor $\alpha\sim 1$ here). The two field
scales introduced in Eqs.~(\ref{field1}) and (\ref{eq:4}) can be
obtained from a comparison [in the argument of logarithmic function
Eq.~(\ref{MR})] of the dephasing term $T/T_{\rm cr}$ with the linear
and quadratic in $B$ terms, respectively. At lowest temperatures,
there is a clear crossover in the $\delta\sigma_{\rm WL}$ vs. $B$
dependence from $\delta\sigma_{\rm WL}\propto\ln B$ to
$\delta\sigma_{\rm WL}\propto 2\ln B$. Note that the crossover occurs
in the 2D regime, where the typical closed path for a coherent
electron motion spans many grains. It is remarkable that even in the
2D regime there is a clear difference in the magnetoresistance of a
granular system from that of a homogeneous film, see Fig.~3.

\begin{figure}
\includegraphics[width=2.5in]{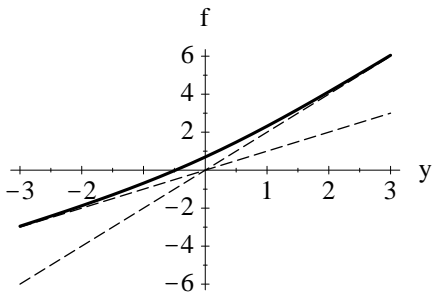}
\caption{Magnetic field dependence (\protect\ref{MR})
of the conductivity at low temperatures, $T \ll T_{\rm cr}
g_{\rm T}/g_{\rm cr}$. The horizontal axis here is proportional to the
logarithm of the applied magnetic field, $y=\ln (Bd^2/\Phi_0) +
\ln (g_{\rm gr}/g_{\rm T})$; the vertical axis is the magnetoresistance
in dimensionless units, $\delta\sigma_{\rm MR} = (e^2/2\hbar)f(y)$
with $f(y) = y + \ln (1 + e^y)$, see Eq.~(\ref{MR}). The crossover between
$\ln B$ and $2\ln B$ dependences is clearly seen.}
\end{figure}

\section{Discussion}

Let us now discuss the temperature dependence of the conductance in various
magnetic fields (Fig.~4). In zero field, the weak localization correction
behaves as $\delta \sigma_{\rm WL}/\sigma_0 \sim g_{\rm T}^{-1}\ln T$
at $T < T_{\rm cr}$ and then crosses over to the power-law behavior,
$\delta \sigma_{\rm WL}/\sigma_0 \sim T_{\rm cr}/(g_{\rm T}T)$, at higher
temperatures. Finite magnetic field
leads to suppression of the WL correction even at the lowest temperature.
Thus, at $B \ll (\Phi_0/d^2)(g_{\rm T}/g_{\rm gr})$
the WL correction becomes temperature-independent. (Note that
$(\Phi_0/d^2)(g_{\rm T}/g_{\rm gr})\ll B_{\phi}^{\rm sg}$.)
At higher temperatures, the dimensionless conductivity
$\delta \sigma_{\rm WL}/\sigma_0$ has the same temperature dependence
as at $B = 0$.
In higher fields, $(\Phi_0/d^2)(g_{\rm T}/g_{\rm gr}) \ll B \ll
B_{\phi}^{\rm sg}$, the same low-temperature saturation occurs at $T =
(Bd^2/\Phi_0)^2 (g_{\rm gr}/g_{\rm T}) T_{\rm cr}$,
see Fig.~4.
\begin{figure}
\label{gbt}
\includegraphics[width=2.5in]{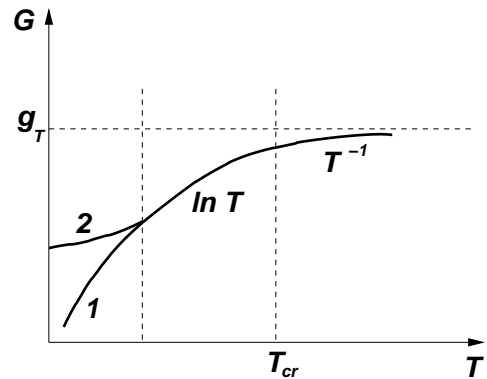}
\caption{Sketch of the temperature dependence of the conductance in
  various magnetic fields: $B=0$ (1); $B \ll B_{\phi}^{\rm sg}$ (2).
  The temperatures at which curve 2
  departs considerably from curve 1 depend on the applied field; these
  temperatures are of the order of $T_{\rm cr} (Bd^2/\Phi_0)$ and
  $T_{\rm cr} (Bd^2/\Phi_0)^2 (g_{\rm gr}/g_{\rm T})$ for $B \ll
  (\Phi_0/d^2)(g_{\rm gr}/g_{\rm T})$ and $B \gg (\Phi_0/d^2)(g_{\rm
    gr}/g_{\rm T})$, respectively.}
\end{figure}
In the highest fields, $B \gg B_{\phi}^{\rm sg}$, the WL correction is
suppressed for all trajectories --- even those lying within a single
grain, and the WL correction disappears at all temperatures.

Note that all magnetic fields which we have discussed above are
too small to change orbital dynamics of electrons. Indeed, the cyclotron
radius, $r_c = mv_Fc/eB$ must be smaller than the mean free path $l$ in
order to affect the electron motion. This corresponds to magnetic fields
$B > (\Phi_0/d^2)(\tau\delta )^{-1}$, with $\tau$ being the momentum
relaxation time in a grain. Since $\tau\delta \ll 1$ (conditions
for metallic diffusive behavior), such fields are well outside
our consideration range.

Apart from the weak localization correction, there is one more
temperature dependent contribution to the conductance --- interaction
correction.  For granular media, it was calculated for all
temperatures in Ref.  \onlinecite{Beloborodov03}. It crosses over from
low- to high-temperature regime at the temperature $g_{\rm T}\delta$,
which is different (much lower) than $T_{\rm cr}$. For a
two-dimensional array, this correction is logarithmic at any
temperatures; for $T \gg g_{\rm T}\delta$, one has $\delta
\sigma/\sigma_0 \sim g_{\rm T}^{-1} \ln (g_{\rm T}E_C/T)$, where $E_C$
is the charging energy in a single grain.
The temperature dependence of the interaction correction is
featureless at  $T\sim T_{\rm cr}$, and therefore it should not mask
the crossover in the temperature dependence of the WL correction.
The interaction correction is also independent of the magnetic field.
Thus it does not affect the crossover in the magnetic field dependence
of the conductance, which is induced by the granular structure.
The measurements of
the conductance therefore can be used to characterize the medium.

Let us finally give some estimates. We consider metallic grains
of a size of $500$ nm, which can be easily produced lythographycally
\cite{Fazio}. They have the level spacing of order $\delta/k_B \sim
20$mK. Choosing $g_{\rm T} = 10$, we
obtain the crossover temperature $T_{\rm cr} = 2$K, that can be easily
observed experimentally.

This work was supported by NSF Grants DMR02-37296, DMR04-39026, and
EIA02-10736 (University of Minnesota), by the U.S. Department of
Energy, Office of Science via the contract No. W-31-109-ENG-38, by the
Minerva Einstein Center (BMBF), and by Transnational Axis Program
RITA-CT-2003-506095 at the Weizmann Institute of Science. We
acknowledge useful discussions with Igor Aleiner, Yuval Gefen, and Alex
Kamenev.

{\em Note added:} After completing this work, we noticed that a formula
  similar to our Eq.~(\ref{MR}) was derived, with a different method,
  in the work by Biagini {\it et al} very recently~\cite{biagini}. We
  are grateful to Andrei Varlamov for the discussion of relation between
  the two works.

%%%%%%%%%%%%%%%%%%%%%%%%%%%%%%%%%%%%%%%%%%%%%%%%%%%%%%%%%%%%%%%%%%%%%%%%%%%%%%%
%%%%%%%%%%%%%%%%%%%%%%%%%%%%%%%%%%%%%%%%%%%%%%%%%%%%%%%%%%%%%%%%%%%%%%%%%%%%%%%


\begin{thebibliography}{99}

\bibitem{Gershenson} B.L. Altshuler, A.G. Aronov, M.E. Gershenson, and
  Yu.V. Sharvin, {\em Quantum effects in disordered metal films}, in:
  Sov. Sci. Rev. A. Phys. {\bf 9}, 223 (1987).

\bibitem{Beenakker} C.~W.~J.~Beenakker and H.~van Houten, Solid State
Physics {\bf 44}, 1 (1991).

\bibitem{Sarachik} E.~Abrahams, S.~V.~Kravchenko, and M.~P.~Sarachik,
Rev. Mod. Phys. {\bf 73}, 251 (2001).

\bibitem{Goldman} B.~G.~Orr, H.~M.~Jaeger, A.~M.~Goldman, and C.~G.~Kuper,
Phys. Rev. Lett. {\bf 56}, 996 (1986).

\bibitem{Dynes}R.~C.~Dynes and J.~P.~Garno,
Phys. Rev. Lett. {\bf 46}, 137 (1981);
R.~C.~Dynes, J.~P.~Garno, G.~B.~Hertel, and T.~P.~Orlando,
Phys. Rev. Lett. {\bf 53}, 2437 (1984).

\bibitem{Fazio} R.~Fazio and H.~S.~J.~van der Zant, Phys. Rep. {\bf 355},
  235 (2001).

\bibitem{Yacoby} S.~Ilani, A.~Yacoby, D.~Mahalu, and H.~Shtrikman,
Phys. Rev. Lett. {\bf 84}, 3133 (2000).

\bibitem{Birge} F.~Pierre, A.~B.~Gougam, A.~Anthore,
  H.~Pothier, D.~Esteve, and N.~O. Birge,  Phys. Rev. B {\bf 68},
  085413 (2003).

\bibitem{AAK} B.~L. Altshuler, A.~G. Aronov, and D.~E. Khmelnitsky, J.
  Phys. C {\bf 15}, 7367 (1982).

\bibitem{Beloborodov01} I.S. Beloborodov, K.B. Efetov, A. Altland, and
  F.W.J. Hekking, Phys. Rev. B {\bf 63}, 115109 (2001).

%\bibitem{Beenakker1} H.~U.~Baranger and P.~A.~Mello, Phys. Rev. Lett.
%  {\bf 73}, 142 (1994).

%\bibitem{Beenakker2} C.~W.~J.~Beenakker, Rev. Mod. Phys. {\bf 69},
%  731 (1997).

\bibitem{Sivan} U.~Sivan, Y.~Imry, and A.~G.~Aronov, Europhys.
 Lett. {\bf 28}, 115 (1994); Ya.~M.~Blanter, Phys. Rev. B
{\bf 54}, 12807 (1996); B.~L.~Altshuler, Y.~Gefen, A.~Kamenev, and
L.~S.~Levitov, Phys. Rev. Lett. {\bf 78}, 2803 (1997).


\bibitem{ABG} I.~L. Aleiner, P.~W. Brouwer, and L.~I. Glazman, Phys. Rep.
  {\bf 358}, 309 (2002).

\bibitem{Houzet} M.~Houzet, D.~A.~Pesin, A.~V.~Andreev, and
L.~I.~Glazman, Phys. Rev. B {\bf 72}, 104507 (2005).

\bibitem{Rammer} G.~D.~Mahan, {\em Many-Particle Physics}, Kluwer, New
York (2000).


\bibitem{AA} B.~L.~Altshuler and A.~G.~Aronov, in {\em
    Electron-Electron Interaction In Disordered Systems}, eds.
    A.~J.~Efros and M.~Pollak, p. 1, North-Holland, Amsterdam (1985).

\bibitem{VA} M.G. Vavilov, I.A. Aleiner, Phys. Rev. B {\bf 60},
  R16311 (1999).

\bibitem{AB02} I.~L.~Aleiner and Ya.~M.~Blanter, Phys. Rev. B {\bf
65}, 115317 (2002).

\bibitem{Beloborodov03} I.~S.~Beloborodov, K.~B.~Efetov, A.~V.~Lopatin,
and V.~M.~Vinokur, Phys. Rev. Lett. {\bf 91}, 246801 (2003).


\bibitem{biagini}  C. Biagini, T. Caneva, V. Tognetti, and
  A.A. Varlamov, Phys. Rev. B. {\bf 72}, 041102 (2005).

\end{thebibliography}
\end{document}